# Pulsed Schlieren Imaging of Ultrasonic Haptics and Levitation using Phased Arrays

Michele Iodice[1]

[1]*Ultrahaptics Ltd, Bristol, United Kingdom*
email: michele.iodice@ultrahaptics.com

William Frier[2]

[2]*University of Sussex, Brighton, United Kingdom*
email: w.frier@sussex.ac.uk

James Wilcox[1], Ben Long[1], Orestis Georgiou[1]

Ultrasonic acoustic fields have recently been used to generate haptic effects on the human skin as well as to levitate small sub-wavelength size particles. Schlieren imaging and background-oriented schlieren techniques can be used for acoustic wave pattern and beam shape visualization. These techniques exploit variations in the refractive index of a propagation medium by applying refractive optics or cross-correlation algorithms of photographs of illuminated background patterns. Here both background-oriented and traditional schlieren systems are used to visualize the regions of the acoustic power involved in creating dynamic haptic sensations and dynamic levitation traps. We demonstrate for the first time the application of background-oriented schlieren for imaging ultrasonic fields in air. We detail our imaging apparatus and present improved algorithms used to visualize these phenomena that we have produced using multiple phased arrays. Moreover, to improve imaging, we leverage an electronically controlled, high-output LED which is pulsed in synchrony with the ultrasonic carrier frequency.

## 1. Introduction

Schlieren imaging (or photography) is an optical process used to visualize density gradients and inhomogeneities (from the German "schliere"; meaning "streak") in the pressure of fluids, that are invisible to the human eye [1, 2]. The application of traditional schlieren imaging ranges from imaging of sound waves [3] to that of inhalation in humans and animals, shock waves and heat waves [2, 4]. The traditional implementation of a schlieren optical setup for acoustic pressure field visualization uses light from a single collimated source illuminating an object. The variations of density gradient in the fluid cause proportional variations in the refractive index which distort the collimated beam of light and create spatial variations of light intensity. The latter can be visualized with an optical system (i.e. a camera) as darker or brighter patches corresponding to positive or negative fluid density gradients. The diffracted light is usually focussed using a pair of optical lenses, or one parabolic mirror. A knife-edge cut-off is positioned at the focal point, to adjust the dynamic range of the system by preventing some of the reflected light from reaching the viewing screen [2, 5-7].

Background-oriented schlieren (BOS) is a type of schlieren imaging technique based on the analysis of image displacements of a patterned background [4, 8-11]. The distortions created by the presence of a fluid pressure in between the optical camera and the background are detected as vertical and horizontal shifts of parts of the background image relative to its original position. Typically, the background is a randomly speckled background, and the shifts of parts of the image with respect to a reference image are estimated via traditional spatial cross-correlation algorithms. They work by subsampling two sequential digital images at one particular area via squared interrogation windows. An average spatial shift of particles may be observed within these image samples, provided a flow of fluid is present in one of the two images. Interrogation windows are usually overlapped for more than half of their length, to tackle aliasing [12]. BOS systems are simpler to construct since they do not require additional optical lenses and mirrors and support unlimited field of view, but they are often less sensitive to gradient changes than the traditional schlieren imaging [2]. BOS systems are traditionally applied to visualization of heated air flows, air vortexes, shock waves and underwater ultrasonic fields [10, 11, 13]. To the authors' knowledge, there are no existing applications of BOS systems to the visualization of ultrasound in air.

Ultrasonic acoustic fields have recently been used to generate haptic effects as well as levitate small sub-wavelength size particles. In haptic feedback and levitation systems, the amplitude and phase angle of the acoustic field is controlled by defining one or more control points in space, in order to create regions of focussed acoustic power, which are amplitude or spatially modulated to create tangible effects, or null points, which serve as acoustic traps for levitating sub-wavelength particles. In this work both effects are achieved using controlled phased arrays, in which the phase of the signal is electronically adjusted. Haptic effects are generated using a one-sided array, whilst levitation is achieved by means of a double-sided phased array in a sandwich arrangement [14-16]. Note that both haptic and levitation effects are created via the focusing of ultrasound waves to a single point in space. For haptics this point is amplitude modulated at 200 Hz as to create vibrations on the skin that are in the range of touch [17], while for levitation the focus is unmodulated. The sandwich arrangement ensures that opposing pressure forces are cancelled, therefore creating a levitation trap. Similar levitation effects could be achieved in many other fashions including a parabolic arrangement of in-phase transducers facing each other and single-sided arrays [18-22].

In this paper, we detail our use of traditional schlieren imaging and BOS techniques to visualize the focussing of the acoustic power that creates dynamic haptic sensations as well as dynamic acoustic traps. A single-sided Ultrahaptics (UHEV1) phased array is used to generate the dynamic focused regions, whilst a modified version (UHEV2) composed of two emitting phased arrays opposed to each other is used to generate the acoustic trap by locally concentrating the synchronized acoustic energy coming from both sides. Since the power used for levitation is comparable to that used for haptics, we modify both schlieren techniques as to amplify their imaging performance. Namely, we leverage a high-output light-emitting diode (LED) which pulses in synchrony with the ultrasonic carrier wave, as the optical source of light for the schlieren imaging techniques. Pulsed schlieren imaging is used by many researchers to increase the sensitivity of the systems to small gradient variations [8, 23-25]. Therefore, exposure time of the camera is defined by the duration of the light pulse, as well as by the shutter speed.

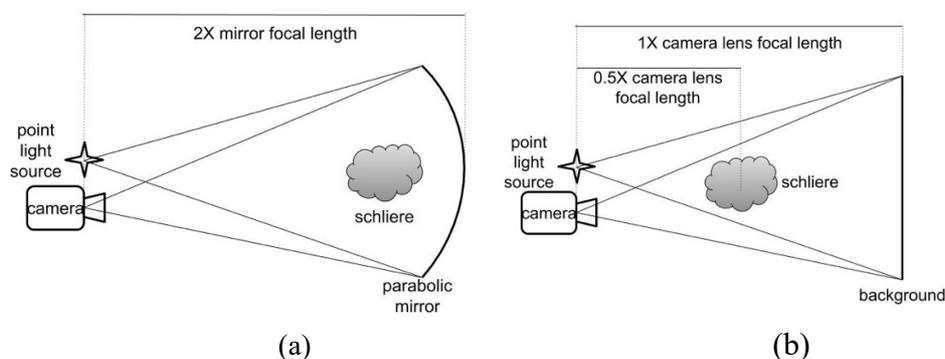

Figure 1: Experimental set-up for traditional schlieren imaging (a) and BOS imaging (b).

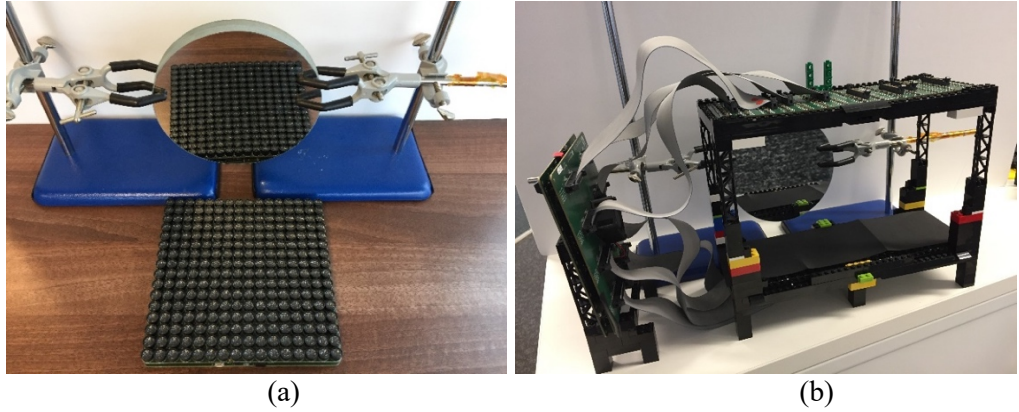

(a)                                        (b)

Figure 2: The one-sided phased array used to control the acoustic field to create haptics effects (a) and the double-sided phased array used for levitating particles (b), with the parabolic mirror immediately behind, following the scheme of Figure 1. It is worth to notice that only the left-end side of the levitation apparatus of (b) was used.

The main contributions of this paper are therefore the first ever use of schlieren imaging techniques to visualize 1) haptic ultrasonic fields, and 2) ultrasonic levitation traps that are dynamic in space, and the first ever use of BOS imaging to visualize 3) the airborne ultrasonic pressure field. An effort has been made to accurately detail the above methods and thus enable reproducibility. We conclude with a discussion on the potential of these methods being used as interactive visualization tools for the real-time imaging of ultrasonic-based haptic feedback and levitation systems.

## 2. Methods and apparatus

The schlieren imaging apparatus consists of a Cassini telescope parabolic mirror, with a focal length of 1300 mm and diameter of 160 mm, a Canon EOS 1200D camera equipped with a Canon EF 70-200mm f/4 L USM zoom lens and a 5184 × 3456 pixels sensor. The BOS imaging uses a 297 mm × 420 mm speckled background produced with a quasi-random scalar generator, with the pixel values corresponding to white and black quantised to either 0 or 1. In both cases, a stroboscopic, high-output, green LED was used to improve imaging quality.

Figure 1 describes the experimental set-up adopted for the schlieren imaging in this paper. For the traditional schlieren imaging (Figure 1 (a)), the parabolic mirror is positioned at twice its focal length from the LED. The camera is mounted on a rigid tripod and positioned so that the diffracted light is refocused in its projection plane. A knife-edge is also positioned at the location where the diffracted light refocuses. In this case the object to be observed should be as close as possible to the parabolic mirror in order to avoid seeing its reflected image as well. For BOS imaging, there is no need for the parabolic mirror. Instead, a speckled background is positioned at a distance equal to the focal length of the camera lens, in order to maximize the resolution and the sensitivity. The sensitivity of the system is dependent on the camera-to-object distance, the camera specifications and the strength of the schliere [4]. Hence, the object to be observed is typically located halfway between the camera and the background (Figure 1 (b)), in order to maximize the refractive image displacements [8]. Two images of the background behind the acoustic field of interest are recorded using a digital camera; one without the acoustic field, and one with it. It is imperative to minimize the camera motion, shutter jitter and in general to avoid any external disturbances and vibrations that could alter the relative position between the different elements. To further increase the sensitivity, the images are collected with very small aperture and very big exposure time. In this paper, they are recorded with ISO 400, f-stop of f/32 and exposure time of 13 seconds. In fact, while the small aperture reduces the sensitivity but improves the focusing on objects at different depths, the long exposure compensates by increasing sensitivity. Figure 3(a) shows the LED and the optical camera used in the experimental investigation.

The acoustic haptic field is controlled using an Ultrahaptics Evaluation Kit (UHEV1), which comprises a two-dimensional array of 256 transducers (16×16) with a known, regular spatial arrangement. Each of them is capable of emitting a monochromatic, 40 kHz sinusoidal wave. Focus points are provided as control inputs to a software solver which determines how to generate an acoustic field to produce them. For acoustic levitation, a modified version of the device (UHEV2) is assembled, using a double-sided phased array of 252 transducers (9×28). They are oriented facing each other by means of a structure made of Lego bricks separated by a distance of 175 mm. Waves are focused onto one or more control inputs from both array plates that destructively interact to form null points surrounded by levees (pressure maxima) thereby acoustically trapping and levitating small sub-wavelength polystyrene particles. Both setups are shown in Figure 2.

Figure 3(b) describes the circuit to control the pulse width modulation (PWM) of the LED: the 40 kHz signal is extracted from the array to get the time stamp, while a host controller USB interface sets the wanted PWM used as the input for the LED driver through a field-programmable gate-array (FPGA). In the following experimental investigations, the LED pulses synchronously with the ultrasonic carrier wave, with a pulse length of 5 μs, and in any case always lower than 25 μs, which equals the pulse width of the ultrasonic carrier frequency emitted by the transducers of the phased array.

Figure 4 describes the numerical implementation of the BOS imaging technique used. The images undergo a preliminary phase in which the schlieren image is transformed to match the pixel intensity values histogram of the original image [26]. Moreover, the schlieren image is automatically registered relative to the original image using feature matching techniques [27]. Subsequently, subsampling of the two images are paired and a complex conjugate multiplication of each corresponding pair of Fourier coefficients is performed. The cross-correlation function is finally obtained by inverse transforming the new set of Fourier coefficients. The displacement vector is recovered by finding the highest value in the two-dimensional array of correlation values [12]. Since the Nyquist's sampling criterion limits the maximum recoverable spatial displacement in any direction, the cross-correlation is usually performed using overlapping windows. In this paper, the displacement vector is recovered using a 16 × 16 pixels interrogation window, with 12 pixels overlap. This leads to a final image with resolution of 1292 × 860 pixels. Since the techniques utilized in the numerical processing involved in BOS imaging are well-known, a rigorous overview is omitted for the sake of brevity.

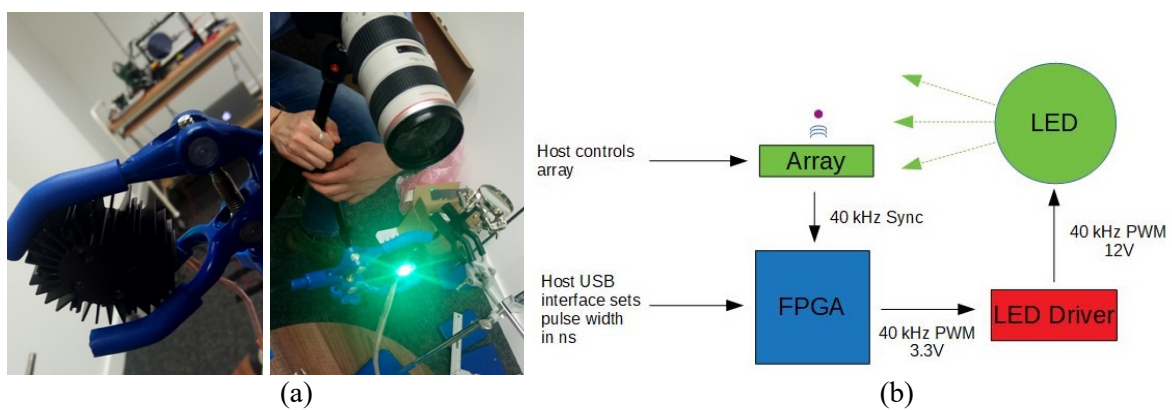

(a)                       (b)

Figure 3: The LED and the optical camera used for the experiment (a) and schematic of the circuit to control the PWM of the LED (b)

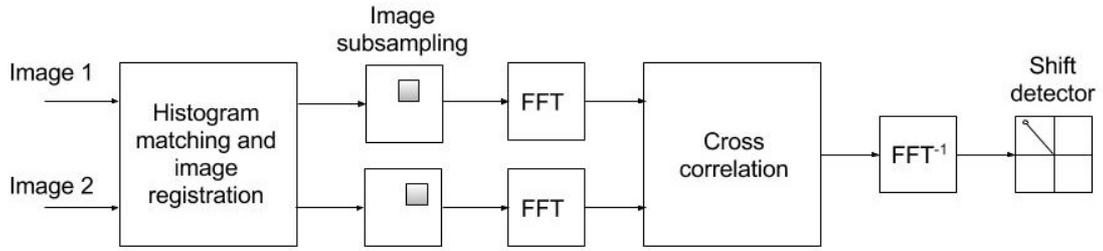

Figure 4: Flow-chart of the numerical processing involved in BOS imaging.

## 3. Results

The results of traditional schlieren imaging to visualize focal points used in haptic feedback systems are shown in Figure 5. The acoustic field is interacting with a cardboard flat sheet in Figure 5(a), and interacting with a hand in Figure 5(b). Photographs are taken with ISO 100, f-stop of f/2.8 and exposure time of 1/1600 seconds. In all the cases, it is possible to notice darker and brighter fringes corresponding to positive and negative density gradient areas of the air medium, caused by the propagation of focussed ultrasonic compressional waves. The distance between two consecutive darker or brighter fringes is equal to half the ultrasonic carrier wavelength, i.e. to 4.3 mm. Reflections caused by the interaction with the cardboard and the hand are also evident in Figure 5, as almost the entirety of the sound energy is believed to be reflected by these surfaces. The variation in density gradients gets stronger as more acoustic power is concentrated closer to the focus point. It is worth noticing the spherical wave-front eventually colliding onto a single focal point near to the reflecting surface or hand. The visualization of a control point following a circular pattern in the plane perpendicular to the phased array is shown in videos here [28, 29], whilst the dynamic interaction of a static focal point with reflective objects is illustrated in this video [30]. The former shows how the wavefield changes, whilst the latter displays the propagation of the compressional wave and the reflection from obstacles.

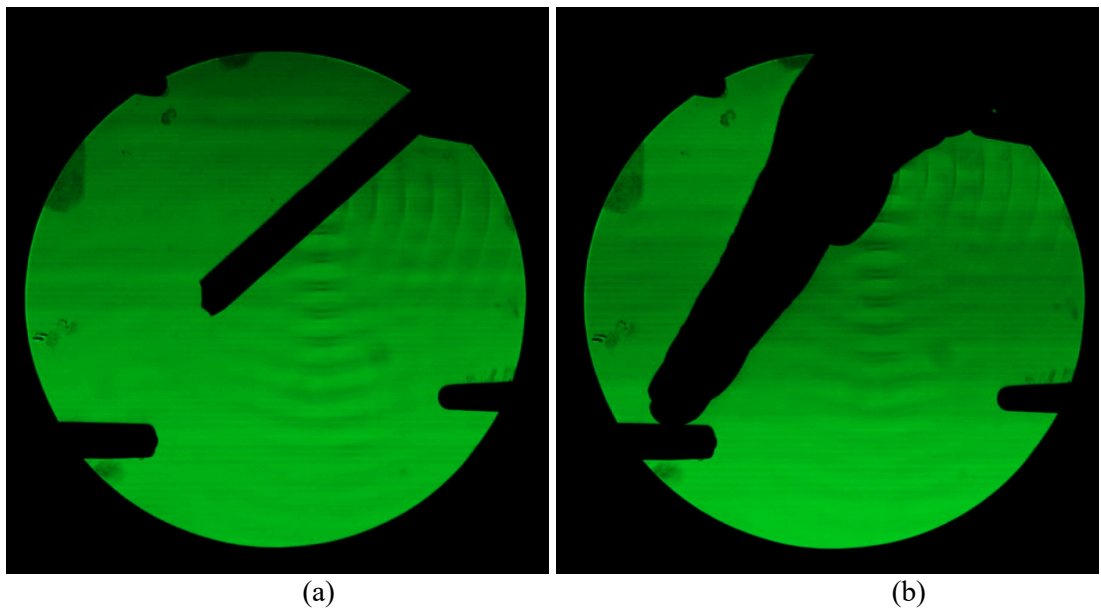

(a)         (b)

Figure 5: The visualization of the pressure field obtained with a single focal point interacting with a flat cardboard sheet (a) and with a hand (b), with traditional schlieren.

The visualization of the acoustic field obtained with the double-sided phased array for objects levitation is shown in Figure 6. A single polystyrene sphere is acoustically levitated in

Figure 6(a), whilst a group made of 5 spheres is levitated in Figure 6(b). They are acoustically trapped in the nodes of the focused waves created by the interaction of the two-opposite facing phased arrays. In both cases, the actual levitation particles and their shadow images are visible in the schlieren photographs. Photographs are taken with ISO speed of 100, f-stop of f/2.8, exposure time of 1/200 sec for the former and exposure time of 1/500 for the latter. Darker and brighter patches are evident and yet again separated by a distance equal to half wavelength. In this case they correspond to the nodes and the antinodes of the standing wave generated from the interaction of the two emitting phased arrays. Polystyrene spheres can be also moved in the 3D space, as shown in videos here [31, 32], as the acoustic field is dynamically controlled adjusting the phase of the signals emitted by each transducer. In these cases, the polystyrene spheres are moved along a circular pattern of 5 cm of diameter in a parallel plane with respect to the double-sided array. It is possible to notice heat waves originating from both sides of the array of transducers due to the electronic and mechanical components warming up.

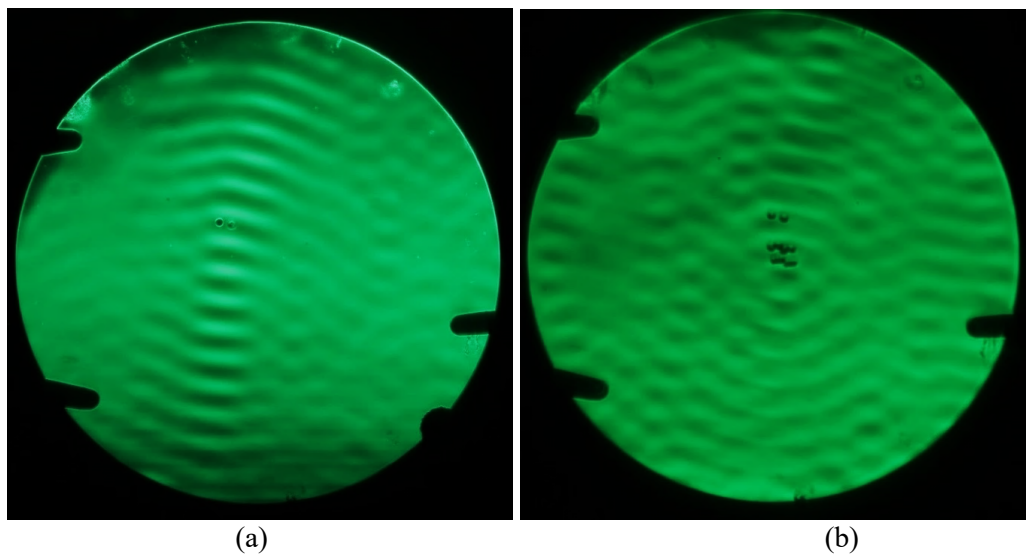

(a) (b)

Figure 6: The visualization of the levitation pressure field of one (a) and of five (b) polystyrene spheres with traditional schlieren.

We apply BOS imaging to visualize the acoustic field produced by the double-sided phased array, following the method and numerical processing described in Section 2. It results in the RGB image presented in Figure 7(a). The unprocessed image with the focussed background and the levitating spheres, is shown in Figure **7**(b). Three levitated spheres are visible near the centre of the image, since they are associated to the biggest displacements and hence to the most vivid colour. They are surrounded by green fringes associated to focussed regions with high pressure density gradients. The latter represent the anti-nodes of the focused standing wave and they are separated by a distance equal to half the carrier wave frequency. Black regions are instead associated to zero relative movements between the original and the schlieren image. Noisy regions in the image are the results of poor illumination or relative movements between the original and the schlieren image. More details are discussed in the next section.

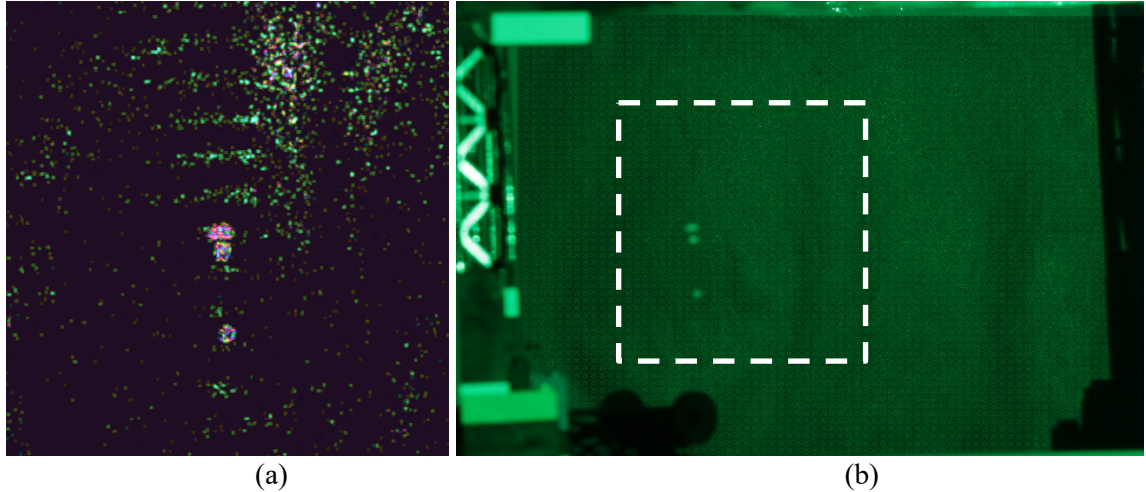

(a)                          (b)

Figure 7: The visualization of the levitation pressure field with BOS imaging (a), and the unprocessed image with the focussed background (b). The dotted rectangle represents the portion of acoustic field imaged in (a).

## 4. Discussion

Traditional pulsed schlieren imaging and BOS have been demonstrated to be good tools for the visualization of ultrasonic pressure fields. The set-ups described in Section 2 were relatively inexpensive (compared to other acoustic field visualization tools, like Laser Doppler Vibrometers (LDV)) and are easy to reproduce. Results shown in Section 3 were consistent and in agreement with the expected results obtained from simulations and from LDV measurements.

The traditional schlieren imaging system described herein has demonstrated the potential of being employed as interactive visualization tools for real-time qualitative and possibly quantitative imaging of ultrasonic-based haptic feedback, or levitation systems. Haptics and levitation could be improved by observing the associated acoustic field and how it interacts in real-time with objects within (hands, particles), using schlieren imaging.

BOS imaging is applied here for the first time to visualize ultrasonic pressure fields in air. BOS imaging is cheaper than traditional schlieren imaging but requires post processing and it does not support interactivity. The dependency of the variations of the refractive index from density makes the application of BOS to airborne ultrasound challenging. Furthermore, BOS imaging is very sensitive to any changes of the relative distance between the different components of the apparatus and to any small vibration caused by the shutter jitter or external artificial or natural sources present in the room. All of them are relevant causes of additional noise. For the visualization of the ultrasonic wavefield, BOS is also very sensitive to the shutter speed. It is thought that longer exposure time would lead to a better visualization of the pressure field.

## 5. Conclusions

In this work we have presented experimental investigations that use pulsed schlieren systems to visualize for the first time the ultrasonic pressure fields involved in creating dynamic haptic sensations and dynamic levitation traps. The acoustic fields were created using different arrangements of commercially available ultrasonic phased arrays. The two schlieren systems employed were further improved by leveraging an electronically controlled LED which pulsed synchronously with the ultrasonic signals.

# Acknowledgements

This project has received funding from the European Union's Horizon 2020 research and innovation programme under grant agreement No 737087.

# REFERENCES

1 Toepler, A.J.I., *Beobachtungen nach einer neuen optischen methode: Ein beitrag experimentalphysik*. 1906: W. Engelmann.
2 Mazumdar, A., *Principles and techniques of schlieren imaging systems*. Columbia University Computer Science Technical Reports, 2013: p. 14.
3 Don-Liyanage, D.K. and D.C. Emmony, *Schlieren imaging of laser-generated ultrasound*. Applied Physics Letters, 2001. **79**(20): p. 3356-3357.
4 Hargather, M.J. and G.S. Settles, *Background-oriented schlieren visualization of heating and ventilation flows: HVAC-BOS*. Hvac&R Research, 2011. **17**(5): p. 771-780.
5 Azuma, T., A. Tomozawa, and S.-i. Umemura, *Observation of ultrasonic wavefronts by synchronous Schlieren imaging*. Japanese journal of applied physics, 2002. **41**(5S): p. 3308.
6 Neumann, T. and H. Ermert, *Schlieren visualization of ultrasonic wave fields with high spatial resolution*. Ultrasonics, 2006. **44**: p. e1561-e1566.
7 Niederhauser, J., et al., *Real-time optoacoustic imaging using a Schlieren transducer*. Applied Physics Letters, 2002. **81**(4): p. 571-573.
8 Raffel, M., *Background-oriented schlieren (BOS) techniques*. Experiments in Fluids, 2015. **56**(3): p. 60.
9 Goldhahn, E. and J. Seume, *The background oriented schlieren technique: sensitivity, accuracy, resolution and application to a three-dimensional density field*. Experiments in Fluids, 2007. **43**(2-3): p. 241-249.
10 Adkin, M.J. and J. Lamb. *Large Field Background Oriented Schlieren for Visualising Heated Air Projection of Fan Heaters*. in *Proceedings of the International Conference on Heat Transfer and Fluid Flow, Prague, Czech Republic*. August 11-12, 2014.
11 Meier, G., *Computerized background-oriented schlieren*. Experiments in fluids, 2002. **33**(1): p. 181-187.
12 Willert, C.E. and M. Gharib, *Digital particle image velocimetry*. Experiments in fluids, 1991. **10**(4): p. 181-193.
13 Pulkkinen, A., J.J. Leskinen, and A. Tiihonen, *Ultrasound field characterization using synthetic schlieren tomography*. The Journal of the Acoustical Society of America, 2017. **141**(6): p. 4600-4609.
14 Carter, T., et al. *UltraHaptics: multi-point mid-air haptic feedback for touch surfaces*. in *Proceedings of the 26th annual ACM symposium on User interface software and technology*. 2013. ACM.
15 Long, B., et al., *Rendering volumetric haptic shapes in mid-air using ultrasound*. ACM Transactions on Graphics (TOG), 2014. **33**(6): p. 181.
16 Gavrilov, L., *The possibility of generating focal regions of complex configurations in application to the problems of stimulation of human receptor structures by focused ultrasound*. Acoustical Physics, 2008. **54**(2): p. 269-278.
17 Obrist, M., S.A. Seah, and S. Subramanian. *Talking about tactile experiences*. in *Proceedings of the SIGCHI Conference on Human Factors in Computing Systems*. 2013. ACM.
18 Marzo, A., A. Barnes, and B.W. Drinkwater, *TinyLev: A multi-emitter single-axis acoustic levitator*. Review of Scientific Instruments, 2017. **88**(8): p. 085105.
19 Norasikin, M.A., et al. *Acoustic levitation by a metamaterial-based cloak*. in *24th International Congress on Sound and Vibration*. 2017. London.

20. Ochiai, Y., T. Hoshi, and I. Suzuki. *Holographic whisper: Rendering audible sound spots in three-dimensional space by focusing ultrasonic waves*. in *Proceedings of the 2017 CHI Conference on Human Factors in Computing Systems*. 2017. ACM.
21. Ochiai, Y., T. Hoshi, and J. Rekimoto, *Pixie dust: graphics generated by levitated and animated objects in computational acoustic-potential field.* ACM Transactions on Graphics (TOG), 2014. **33**(4): p. 85.
22. Andrade, M.A., et al., *Acoustic levitation of an object larger than the acoustic wavelength.* The Journal of the Acoustical Society of America, 2017. **141**(6): p. 4148-4154.
23. Hanafy, A. and C. Zanelli. *Quantitative real-time pulsed Schlieren imaging of ultrasonic waves*. in *Ultrasonics Symposium, 1991. Proceedings., IEEE 1991*. 1991. IEEE.
24. Bencs, P., et al. *Synchronization of particle image velocimetry and background oriented schlieren measurement techniques*. in *Proc. 8th Pacific Symposium on Flow Visualization and Image Processing*. 2011.
25. Hernandez, R., et al., *Visualization and Computation of Quantified Density Data of the Rotor Blade Tip Vortex*, in *New Results in Numerical and Experimental Fluid Mechanics VIII*. 2013, Springer. p. 321-329.
26. Gonzalez Rafael, C., E. Woods Richard, and L. Eddins Steven, *Digital image processing using MATLAB.* Editorial Pearson-Prentice Hall. USA, 2004.
27. Evangelidis, G., *IAT: A Matlab toolbox for image alignment*. 2013.
28. Ultrahaptics. *Pulsed schlieren imaging of ultrasonic haptics and levitation using phased arrays [video file1]*. 2018; Available from: www.youtube.com/watch?v=WuzGLxnWqDs.
29. Ultrahaptics. *Pulsed schlieren imaging of ultrasonic haptics and levitation using phased arrays [video file2]*. 2018; Available from: www.youtube.com/watch?v=3V8zBuZS7uc.
30. Ultrahaptics. *Pulsed schlieren imaging of ultrasonic haptics and levitation using phased arrays [video file3]*. 2018; Available from: www.youtube.com/watch?v=jTMu2rmdako.
31. Ultrahaptics. *Pulsed schlieren imaging of ultrasonic haptics and levitation using phased arrays [video file4]*. 2018; Available from: www.youtube.com/watch?v=TH0YaYajs48.
32. Ultrahaptics. *Pulsed schlieren imaging of ultrasonic haptics and levitation using phased arrays [video file5]*. 2018; Available from: www.youtube.com/watch?v=KdmZV-4ff0M.